\documentclass{sigchi}
\usepackage{siunitx}
\usepackage{float}
\usepackage[show]{chato-notes} 
\usepackage{amsmath}
\usepackage[titlenumbered,ruled]{algorithm2e}
\usepackage{algpseudocode}
\usepackage{pifont}
\usepackage{times}
\usepackage{url}
\usepackage{tabulary}
\usepackage{booktabs}
\usepackage{multirow}
\usepackage{epstopdf}
\usepackage{tablefootnote}
\usepackage{footnote}
\usepackage{balance}  
\usepackage{graphics} 
\usepackage[T1]{fontenc}
\usepackage{txfonts}
\usepackage{mathptmx}
\usepackage[pdftex]{hyperref}
\usepackage{color}
\usepackage{textcomp}
\usepackage{subfigure}
\usepackage{microtype} 
\usepackage{ccicons}  
\usepackage[utf8]{inputenc} 

\newcounter{NumberOfComments}
\stepcounter{NumberOfComments}

\newcounter{JSNumberOfComments}
\stepcounter{JSNumberOfComments}

\newcounter{JRNumberOfComments}
\stepcounter{JRNumberOfComments}

\def\plaintitle{}

\def\emptyauthor{}
\def\plainkeywords{}

\makeatletter
\def\url@leostyle{%
  \@ifundefined{selectfont}{
    \def\UrlFont{\sf}
  }{
    \def\UrlFont{\small\bf\ttfamily}
  }}
\makeatother
\def\pprw{8.5in}
\def\pprh{11in}

\definecolor{linkColor}{RGB}{6,125,233}
\hypersetup{%
  pdftitle={\plaintitle},
  pdfauthor={\emptyauthor},
  pdfkeywords={\plainkeywords},
  bookmarksnumbered,
  pdfstartview={FitH},
  colorlinks,
  citecolor=black,
  filecolor=black,
  linkcolor=black,
  urlcolor=linkColor,
  breaklinks=true,
}

\begin{document}

\title{Demographics of News Sharing in the U.S. Twittersphere%
\titlenote{\textcolor{red}{\textbf{This is a pre-print of a paper accepted to appear at HyperText 2017}}.\\ \\
\textcolor{blue}{Demographics of News Sharing in the U.S. Twittersphere.\\
Julio C. S. Reis, Haewoon Kwak, Jisun An, Johnnatan Messias, and Fabrício Benevenuto\\
In Proceedings of the 28th ACM Conference on Hypertext and Social Media (HT'17). Prague, Czech Republic. July 2017.\\\\
\textbf{Bibtex: \url{http://johnnatan.me/bibtex/reis_ht_2017.bib}\\
For more published work: \url{http://homepages.dcc.ufmg.br/~julio.reis/}}}}}

\numberofauthors{4}
\author{%
\alignauthor
Julio C. S. Reis\\
       \affaddr{Universidade Federal de Minas Gerais}\\
       \affaddr{Belo Horizonte, Brazil}\\
       \email{julio.reis@dcc.ufmg.br}
\alignauthor
Haewoon Kwak\\
       \affaddr{Qatar Computing Research Institute}\\
       \affaddr{Doha, Qatar}\\
       \email{haewoon@acm.org}
\alignauthor
Jisun An\\
       \affaddr{Qatar Computing Research Institute}\\
       \affaddr{Doha, Qatar}\\
       \email{jan@hbku.edu.qa}
\alignauthor
Johnnatan Messias\\
       \affaddr{Universidade Federal de Minas Gerais}\\
       \affaddr{Belo Horizonte, Brazil}\\
       \email{johnnatan@dcc.ufmg.br}
\alignauthor
Fabricio Benevenuto\\
       \affaddr{Universidade Federal de Minas Gerais}\\
       \affaddr{Belo Horizonte, Brazil}\\
       \email{fabricio@dcc.ufmg.br}
}

\maketitle

\begin{abstract}

The widespread adoption and dissemination of online news through social media systems have been revolutionizing many segments of our society and ultimately our daily lives. In these systems, users can play a central role as they share content to their friends. Despite that, little is known about news spreaders in social media. In this paper, we provide the first of its kind in-depth characterization of news spreaders in social media. In particular, we investigate their demographics, what kind of content they share, and the audience they reach. Among our main findings, we show that males and white users tend to be more active in terms of sharing news, biasing the news audience to the interests of these demographic groups.  Our results also quantify differences in interests of news sharing across demographics, which has implications for personalized news digests.

\end{abstract}


\section{Introduction}

In recent years, with the huge success of Twitter and Facebook, social media has become one of the most important channels in news diffusion. In particular, Twitter's unique concepts of asymmetric ``follow'' and ``retweet'', which were later adopted by Facebook, allow users to follow each other's updates and propagate interesting pieces of information quickly and broadly~\cite{kwak2010twitter}.  Such great power to disseminate information embedded in social media naturally has attracted the news media.  As a result, a majority of U.S. adults (62\%)  get news mostly on social media, according to a new survey by Pew Research Center~\cite{news-social-media}.

Along with their traditional channels, news media manage their presence in social media by creating Twitter accounts and publishing tweets containing URLs that link their news media sites.  For those accounts, it is clearly visible who the audience is -- their followers. Furthermore, as any Twitter user can share URLs to news media web sites, Twitter users exposed to news media's tweets through retweets can also be visible and accounted as audience.  We call these users \textit{news spreaders} in the rest of this paper.  This form of sharing of news URLs has long been a pervasive practice in social media, but its role and impact are relatively unexplored.

In this work, we characterize news spreaders in Twitter along three dimensions: 1) their demographics (who they are), 2) their news shared (what they share), and 3) their impact (why they are important).  To this end, the inference of demographics of Twitter users is essential.  Among various techniques that have been proposed~\cite{mislove2011understanding}, we use state-of-the-art techniques to locate Twitter users and infer their demographics based on profile photos.  

Through a longitudinal data collection of news spreaders and their URL sharing behavior of five popular global news media, we test how similar news URL sharing is to typical URL sharing in terms of demographics of spreaders.  We find a statistically significant trend that white males participate more in news URL sharing than other  race-gender groups.  This suggests that news spreaders have unique characteristics, which cannot be easily perceived for typical URL spreaders in Twitter. Thus, our work is essential to understand news spreaders correctly.  

We then answer the above research questions. First, we examine demographics of news spreaders. By comparing the followers of news media accounts, we discover huge differences in terms of race-gender demographics.  This suggests that we need to have a broader definition of the exposure of the news media on social media that are not only a set of followers~\cite{an2011media} but also news spreaders. 
Second, we examine what kinds of news are shared by news spreaders.  The properties of the pieces of news are defined along three dimensions: topics, author's (journalist's) gender and race, and linguistic analysis~\cite{pennebaker2001linguistic} of news headlines.  These three dimensions have been discovered as important factors in news reading/sharing behavior~\cite{reis2015breaking,shah2015bad}.  
Finally, we answer how important news spreaders are for news media from the perspective of audience expansion: 1) about 59\% of news spreaders do not follow news media accounts in Twitter; 2) the audience brought by the spreaders is much bigger than that of the original followers of the news media; 3) in addition to that the demographics of the spreaders and those of the followers are quite different, the followers of the spreaders are also substantially different from the followers of the news sources in terms of demographics. In other words, the spreaders play an important role in expanding the audience of news in Twitter, which would otherwise be very limited. Lastly, we find that the demographics of news spreaders are related to the popularity of news.  

Our contributions are three-fold: 1) by using a combination of state-of-the-art techniques, we investigate in details aspects of the audience of news media in Twitter, which has been considered as in-house data so far; 2) we suggest a robust statistical framework to test the news URL sharing behavior by comparing it with typical URL sharing behavior;  and 3) Our findings show that news media should understand spreaders and their followers to capture the complete picture of their presence in news media.  News media's direct followers are only the tip of the iceberg of their audience in Twitter in terms of volume and demographics.

The rest of the paper is organized as follows. Section 2 briefly surveys related efforts. Then, we present our experiment methodology and the data gathered. The next three sections cover our results. We conclude the paper by discussing implications from our findings as well as presenting directions for future work.

\section{Related Work}

In this Section, we review existing work related to news sharing along two main dimensions.

\subsection{News Sharing and Propagation} 

Social media services have made personal contacts and relationships more visible and quantifiable than ever before. Users interact by following each others' updates and passing along interesting pieces of information to their friends. This kind of word-of-mouth propagation occurs whenever a user forwards a piece of information to her friends, making users a key element in this process. Not surprisingly, a number of efforts have attempted to quantify and characterize information spread in social networks as well as the role users play in such propagation~\cite{rodrigues@imc11,cha2012world,wu2011says,romero2011influence,icwsm10cha}. For example, Rodrigues et al.~\cite{rodrigues@imc11} showed that retweets are responsible for increasing the audience of URLs by about 2 orders of magnitude. As social media became an important channel in news diffusion, some recent research efforts attempted to investigate how news are shared in these systems. Next, we detail a few approaches that provides news sharing and propagation.


Naveed et al.~\cite{naveed2011bad} showed that bad news tends to spread faster in systems like Twitter. In this same year, also with the use of this same social media, Armstrong et al.~\cite{armstrong2011gender} analyzed how online media companies employ men and women in Twitter feeds and how it connects to portrayals in news. In particular, the authors looked at how mentions of men and women on Twitter may influence mentions in news stories (e.g. newspaper, television). Through the content analysis of newspaper and television tweets at different granularity (i.e. local, regional and national), they found that male mentions were more likely to appear in national news than in regional or local news and more often than female mentions in the print media than on television.


A recent effort~\cite{bright2016social} has tackled the question ``Why are some news articles shared more than others?''. They showed that story importance cues are relevant in driving social sharing and that certain topics (i.e. stories about politics, accidents, disasters, and crime) were less shared. Some topics can be shared in order to improve the users' reputation. This dynamic media attention has inspired other recent studies \cite{whatgetsJisun2017icwsm}. Bright et al.~\cite{bright2016social}, compare different social networks platforms and showed that some kind of news are shared more in one network than the others (e.g. economy news on LinkedIn). 

Unlike previous works, our effort focuses on understanding the dynamics of news sharing on Twitter of each demographic group. Thus, to the best of our knowledge, this is the first effort that investigates intersection between news sharing and demographic information of users, including how these aspects are related.


\subsection{Demographics in Social Media} 


Mislove et al.~\cite{mislove2011understanding} was one of the first researchers that analyze demographic characteristics of Twitter users considering a geographical perspective (i.e. how the demographics vary across different U.S states). After that, several efforts have arisen that investigate demographic information, in various social media, using different strategies for distinct purposes \cite{blevins2015jane, karimi2016inferring, burger2011discriminating}. Particularly, researchers are jointly applying computer science and statistical techniques to support sociological studies using large-scale social media datasets. These studies can range from a simple characterization of to the investigation of more complex causes, including to raising attention to the different levels at which gender biases can manifest themselves on the web~\cite{wagner2015s}.

In \cite{cunha2012gender} the authors used Twitter data to analyze the difference between men and women behavior in terms of dynamics in free tagging environments. The results obtained present gender distinctions in the use of Twitter hashtags, emphasizing it as a social factor influencing the user's choice of specific hashtags on a specific topic. Still about tags (or hashtags), recently, the work presented in \cite{an2016greysanatomy} explored their use by different demographic groups. The demographic characteristics of each user were obtained using \textit{Face++} and the Twitter user's profile picture. The results showed that, although there are more popular hashtags that are commonly used, there are also many group-specific hashtags with non-negligible popularity. Besides that, the researchers show that the strategy of getting demographic data from \textit{Face++} is reliable and provides accurate demographic information for gender and race, encouraging the application of this strategy in other recent efforts~\cite{chakraborty2017icwsm}. We use a similar strategy to gather demographic information.

Nilizadeh et al.~\cite{nilizadeh2016twitter} explore gender inequalities in Twitter, showing that  gender may allow inequality to persist in terms of online visibility. Looking at Pinterest, Gilbert et al.~\cite{gilbert2013need} investigated what role gender plays in the website's social connections. The results highlight a major difference between female and male users regarding their motivations for using this social media. They found that being female means more repins (i.e., more shared content), but fewer followers in comparison with Twitter. Gender differences has also been explored in terms of social media disclosures of mental illness~\cite{de2017gender}.

More recently, An et al.~\cite{an2016multidimensional} examined the news consumption in South Korea (from Daum News portal). The authors analyzed on a large scale the differences in news consumption from a demographic perspective. Through a multidimensional analysis of gender and age differences in news consumption, they quantify such differences along four distinct dimensions: actual news items, topic, issue, and angle. The top 30 news items for each gender and age group in Daum News were used and the demographics information were obtained through the website itself. Overall, focus mainly on quantifying and explaining  differences in news consumption.

More broadly, most of the previous efforts attempt to quantify differences in gender behavior and inequalities in different social media or news systems. Our effort is the first of its kind to provide a characterization of news sharing across different demographic groups. Thus our effort is complementarity to the existing ones.

\begin{table}[t] \centering 
 \small
\begin{tabular} {|l|rrcc|}
\hline
 News Media & \#Shares & \#Authors & Screen name & \#Followers \cr \hline
\textsf{New York Times}   &  14,505 &  1,165  & @nytimes &  1,141\cr
\textsf{Reuters}    & 4,712 & 485 & @Reuters & 1,259\cr
\textsf{The Guardian}    & 4,457 & 844 & @guardian & 1,620\cr 
\textsf{Wall Street Journal} &  1,379 & 313 & @WSJ &  1,445\cr
\textsf{BBC News} &  1,144 & 190 & @BBCBreaking &  1,130  \cr \hline
\end{tabular} 
\caption{Data collection by news source.}
\vspace{-3mm}
\label{tab:dataset}
\end{table}

\section{Methodology}

In order to understand demographics of news sharing in Twitter, first we define our strategy for data collection. Then, we define our strategies for inference of demographic information of each individual Twitter user and collection of information such as category and authors of the news, and followers of each of the news media on Twitter. Our ultimate goal, in this section, consists of reporting our baseline for comparison in order to verify the statistical significance of the results. Next, we briefly describe the methodology adopted for this work, including a discussion of its main limitations.

\subsection{Gathering Twitter}

For this work, we gathered the 1\% random sample of all tweets, through the Twitter Streaming API \footnote{\url{https://dev.twitter.com/streaming/public}}, along a 3 months period, from July to September, 2016. Specifically, we considered only tweets (and retweets) that contain at least one URL and have been shared by U.S. users. We understand that users who share URLs may present a slight difference in behavior compared to others, so, considering our research objective, we only select this set of American users. Besides that, as we are interested in analyzing demographic characteristics, it is important to study users from the same place. For this reason, we consider only U.S. users, filtered by timezone. In total, we retrieved 11,790,679 tweets posted by 11,770,273 U.S. users. From this initial dataset, we infer demographics information about users and build: (i) our news sharing dataset, used in the execution of our experiments, and (ii) our baseline dataset.

\subsection{Inferring Demographics Information} \label{sec:inferring_demographic}

In the literature, several studies present strategies for inference of gender, race, and age. Some efforts attempt to infer the gender of a user from her name \cite{karimi2016inferring, liu2013s, mislove2011understanding}, or the age from Twitter profile descriptions \cite{sloan2015tweets}, by using patterns like `\textit{like 25 yr old}' or `\textit{born in 1990}'. However, in some cases the number of unrealized inferences (e.g. for lack of information) is high (Liu et al.~\cite{liu2013s} reported 66\% users in their dataset did not have a proper name).

To overcome such limitation, in this work, we use the profile picture's URLs of all users in our dataset and use the \textit{Face++} API \footnote{\url{https://www.faceplusplus.com}}, a face recognition platform based on deep learning \cite{yin2016learning}, 
to infer the gender (i.e., male or female), race (limited to Asian, Black\footnote{We called \textit{African-American (AF-AM)} in the rest of this paper.}, and White) and age information from the recognized faces in the profile images. We discarded  users whose profile pictures do not have a recognizable face or have more than one face, according to \textit{Face++}. Our final dataset contains 937,308 unique users located in U.S. with identified demographic information, which are gender, race, and age by \textit{Face++}.

\subsection{Shared News Dataset} \label{sec:news_dataset}

To focus on news sharing in Twitter, we filtered only tweets that shared news URLs from important and different news sources (i.e. BBC News\footnote{\url{http://www.bbc.com}}, The New York Times\footnote{\url{http://www.nytimes.com}}, Reuters Online\footnote{\url{http://www.reuters.com}}, The Wall Street Journal\footnote{\url{http://www.wsj.com/}}, The Guardian\footnote{\url{https://www.theguardian.com}} and BBC News\footnote{\url{http://www.bbc.com}}), known worldwide. All these news sites appear among the most popular ones in the world, according to Alexa.com.\footnote{\url{http://www.alexa.com/topsites/category/News}}  Simultaneously, we gathered information from users who posted each of the tweets including demographic information from \textit{Face++}, as detailed above. From news URLs, we crawled information about them including, title, text, principal image (link - when there is one), authors (when there is one) and date. Lastly, Table~\ref{tab:dataset} shows the dataset used in this work containing 26,211 unique news articles shared by 16,382 unique users. We note that The New York Times is the most widely shared news media in Twitter, in comparison with those news sites considered. Table~\ref{tab:demogr_spreaders} shows the demographic decomposition of those 16,382 users who shared news URLs.

\begin{table}[]
\centering
\begin{tabular}{|c|cc|c|}
\hline
\multirow{2}{*}{Race (\%)} & \multicolumn{2}{c|}{Gender (\%)}        & \multirow{2}{*}{\textbf{Total:}} \\ \cline{2-3}
                      & \multicolumn{1}{c|}{Male} & Female &                        \\ \hline
Asian                 & 5.29                      & 6.05  & 11.34                  \\
AF-AM                 & 6.09                      & 3.80   & 9.89                  \\
White                 & 43.46                     & 35.31  & 78.77                 \\ \hline
\textbf{Total:}       & 54.84                     & 45.16  & 100.00                 \\ \hline
\end{tabular}
\caption{Demographic distribution of news spreaders.}
\vspace{-3mm}
\label{tab:demogr_spreaders}
\end{table}

\subsection{Inferring News Category} \label{sec:infering_category}

\begin{figure}[h!]
	\centering
		\includegraphics[width=0.45\textwidth]{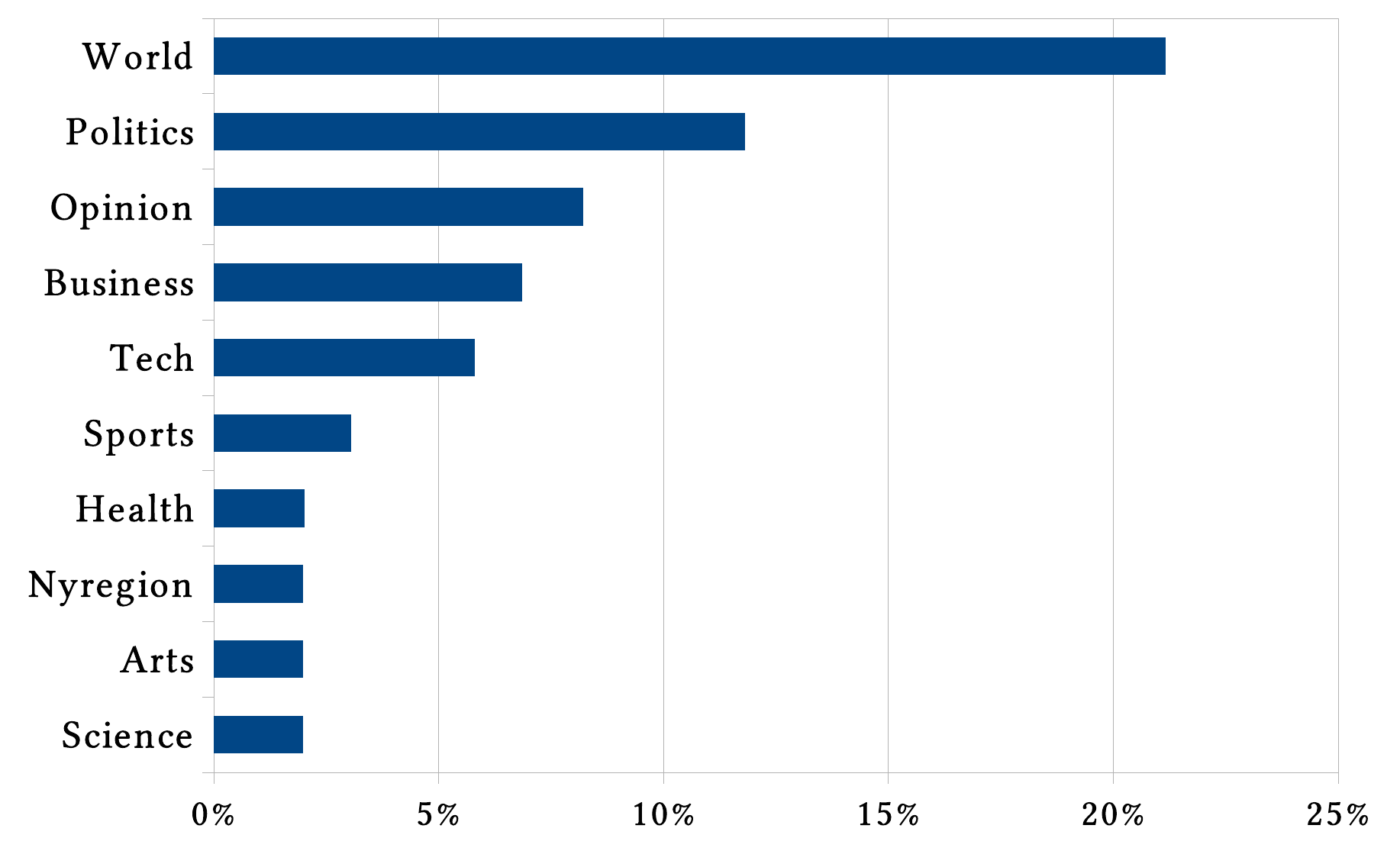}
	\caption{Top 10 most significant news categories.}
	\vspace{-2mm}
	\label{fig:category_top10}
\end{figure}

In order to infer the categories of the news articles, we use meta information embedded in the news URLs.  News media usually have several news sections, such as Politics, Sports, or World News, and group their news articles by these sections. By looking at which section a news article belongs to, we can infer a topical category of the news articles. The section information is often embedded in news URL. For example, the URL \url{http://www.nytimes.com/2016/07/02/us/politics/loretta-lynch-hillary-clinton-email-server.html} represents that the news article is about ``Politics".
We parsed all News URLs and extracted the topic information. The New York Times, The Guardian, and BBC adopt the above mentioned strategy for their URLs, and thus we simply parse their URL and infer the topic of a given news article. 
Reuters and The Wall Street Journal do not have category information in their URLs, however, the news articles have the category information. Thus, we collected news articles and extracted category information by parsing HTML files. We successfully inferred the topical categories of 93.3\% (24,466) of news articles. Figure~\ref{fig:category_top10} shows the proportion of the top 10 most significant news categories. We find that ``World" is the most ``shared'' category (21.16\%), similar to the results in \cite{reis2015breaking}.

\subsection{Finding Journalists in Twitter}

\begin{figure}[!h]
	\centering
		\includegraphics[width=0.48\textwidth]{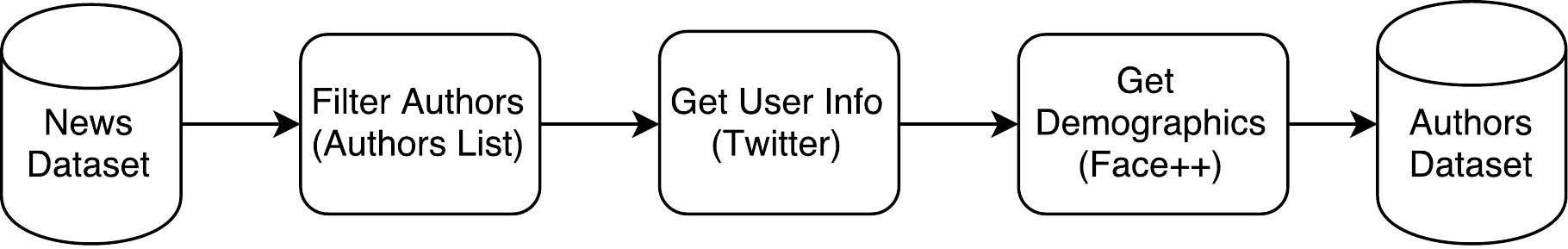}
	\caption{Strategy for collecting news authors.}
	\vspace{-2mm}
	\label{fig:authors_methodology}
\end{figure}

We aim to collect demographics of the authors of news articles in our dataset. Figure~\ref{fig:authors_methodology} shows the procedure for creating an author dataset. 
For each news URL, we collect its title, text, principal image, authors, and date by parsing the original web page.
Then, we search and collect the Twitter profiles of the authors if they have Twitter accounts.  
Then, we infer those authors' demographic characteristics using \textit{Face++} (see Section~\ref{sec:inferring_demographic}).
Table \ref{tab:dataset} shows the number of authors for each news media.
As expected, the largest number of names of distinguished authors we have gathered are from the The New York Times news media, which had the largest number of news shared in Twitter in our dataset.

\subsection{Collecting Followers of News Media in Twitter}

For each news source, we collected their followers in Twitter. Again, we infer their demographics by \textit{Face++}. Table \ref{tab:dataset} presents the total of gathered news media followers in Twitter, including the screen name used for collection. On average, we retrieved 1,319 followers by news source.

\subsection{Baseline Dataset} \label{sec:baseline}

A null model is widely used to estimate the statistical significance of the observed trend  in given data.  
As the null model is randomly generated data that preserve some properties of the original data  (e.g., the degree distribution in complex networks), the same trend  observed from the null model captures its occurrence by chance.
Then, by comparing the trend in the original data with that in the null model, the statistical significance of the observed trend in the original data can be measured.
Table \ref{tab:population_baseline} shows the breakdown of ethnicity and gender of the $\approx 1$  million users who shared URLs in Twitter between July and September 2016.  
We present a detailed description of the comparison with null models.



\begin{table}[]
\centering
\begin{tabular}{|c|cc|c|}
\hline
\multirow{2}{*}{Race (\%)} & \multicolumn{2}{c|}{Gender (\%)}        & \multirow{2}{*}{\textbf{Total:}} \\ \cline{2-3}
                      & \multicolumn{1}{c|}{Male} & Female &                        \\ \hline
Asian                 & 7.07                      & 10.33  & 17.40                  \\
AF-AM                 & 8.52                      & 6.93   & 15.45                  \\
White                 & 31.97                     & 35.18  & 67.15                  \\ \hline
\textbf{Total:}       & 47.56                     & 52.44  & 100.00                 \\ \hline
\end{tabular}
\caption{Demographic distribution of users in the Baseline dataset.}
\vspace{-5mm}
\label{tab:population_baseline}
\end{table}


In this work, whenever we report the number of users with certain properties who share URLs on particular news media, we report \textit{Z}-score by comparing the number of those users in the actual data with that in null models.

Consider that we are interested in users who are Asian and share BBC News. In this case, we denote by $|U_{BBC}|$ the number of users who share BBC News and $|U_{BBC}^{Asian}|$ by the number of Asian among them.  To construct a null model, we create $k$ random samples from a separate huge set of users, which is called Population, where each sample has exactly $|U_{BBC}|$ users.  The demographic information of users in Population is inferred by \textit{Face++}.  For each sample, we count how many Asians are included, $|S_{BBC}^{Asian}|$.  Then, the $Z_{BBC}^{Asian}$ is computed as following:

\begin{equation}
    Z_{BBC}^{Asian}=\frac{|U_{BBC}^{Asian}|-mean(|S_{BBC}^{Asian}|)}{std(|S_{BBC}^{Asian}|)}
\end{equation}

where $mean(\cdot)$ is the mean  and $std(\cdot)$ is the standard deviation of the values from multiple samples.  Intuitively, when the absolute value of $Z$ value becomes bigger (either positive or negative), the trend (more number or less number, respectively) is less likely observed by chance.   In this work, the size of Population is $\approx 1$ million, and $k$=100.  



\subsection{Potential Limitations}

There are a few limitations of our data, discussed next. \\

\noindent
\textbf{Accuracy of the inference by \textit{Face++}}.
First, \textit{(i)} we are limited by accuracy of \textit{Face++} in the inference. \textit{Face++} itself returns confidence levels for the inferred gender and race attributes and returns an error range for inferred age. In our data, the average confidence level reported by \textit{Face++} is 95.24 $\pm$ 0.020\% for gender and 86.12 $\pm$ 0.032\% for race, with a confidence interval of 95\%. Besides that, as the performance of deep learning systems continues to improve, the inferred demographic attributes should become more accurate. Also, recent efforts have used \textit{Face++} for similar tasks and reported high confidence in manual inspections of small samples \cite{an2016greysanatomy, zagheni2014inferring}; Another limitation, is that \textit{(ii)} \textit{Face++} reports race of recognizable faces from images but not the \textit{ethnicity} (e.g. \textit{Hispanic}); Finally, though \textit{(iii)} we had discarded about 70\% of the crawled users (i.e. those users whose profile pictures do not have a recognizable face or have profile pictures in which \textit{Face++} recognized with low confidence). However, we note that the remaining final dataset is still representative and we only provide results that are statistically significant based on well known statistical tests. \\

\noindent
\textbf{Data}.
\textit{(iv)} Our approach to identify users in U.S. may contain users located in the same time zone, but not in the U.S. We, however, believe that these users represent a small fraction of the users, given the predominance of active U.S. users in Twitter \cite{cheng2009inside}; \textit{(v)} We are using the 1\% random sample off all tweets.  Although the 1\% random sample is not the best data to capture all the dynamics happening in Twitter, its limitations are known~\cite{morstatter2014biased} and it is the best available option at our disposal. \\




Even with limitations, we believe that our dataset and methods can provide interesting insights on demographics and news sharing behaviors. In the following sections, we present and discuss the main results from characterizing news spreaders in Twitter along three dimensions: 1) their demographics (who they are), 2) their news shared (what they share), and 3) their impact (why they are important).


\section{Who are the news spreaders?}
Our first research question is to understand who the spreaders are. We compare the demographics of news spreaders with 1) the spreaders of typical URLs in Twitter and 2) the Twitter followers of news media to see whether and to what extent they differ.

\subsection{Typical URL Sharing Vs. News Sharing}

Table~\ref{tab:shares_newspaper} shows, for each news media, the proportion of news URL shares by different demographic groups. For example, for The New York Times, 54.1\% of news shares are made by men and 79.2\% of news shares are by Whites. The numbers in the parenthesis correspond to the Z-values, detailed in Section~\ref{sec:baseline}. We note that the Z-value indicates how news URL sharing behavior is similar or dissimilar from typical URL sharing behavior in terms of demographic composition.

\begin{table}[htbp]
\centering
\frenchspacing
\scriptsize
\begin{tabular}{|c|c|cc|c|}
\hline
\multirow{2}{*}{News media}          & \multirow{2}{*}{Race (\%)} & \multicolumn{2}{c|}{Gender (\%)}  & \multirow{2}{*}{\textbf{Total:}} \\ \cline{3-4}
                                    &                       & \multicolumn{1}{c|}{Male} & Female &                         \\ \hline
\multirow{4}{*}{\textsf{The New York Times}} & Asian        & 5.1 (-9.22)       & 5.9 (-18.02)  &  11.0 (-19.96)                       \\ 
                                    & AF-AM                 & 6.1 (-13.95)      & 3.7 (-15.01)  &  9.8 (-21.75)                        \\
                                    & White                 & 42.8 (26.24)      & 36.4 (2.86)   &  79.2 (31.32)                \\ \cline{2-5} 
                                    & \textbf{Total:}       & 54.1 (15.30)      & 45.9 (-15.30)  &  100.0                           \\ \hline
\multirow{4}{*}{\textsf{Reuters}}   & Asian                 & 3.6 (-8.06)      & 6.8 (-7.62)   &  10.4 (-12.09)                       \\
                                    & AF-AM                 & 7.3 (-3.02)      & 3.7 (-8.70)   &  10.9 (-9.03)                     \\
                                    & White                 & 47.0 (23.21)      & 31.7 (-4.89)  &  78.7 (16.38)                \\ \cline{2-5} 
                                    & \textbf{Total:}       & 57.9 (14.00)      & 42.1 (-14.00)  &  100.0                       \\ \hline
\multirow{4}{*}{\textsf{The Guardian}} & Asian              & 4.9 (-6.11)      & 5.9 (-9.75)   &  10.7 (-12.75)                     \\
                                    & AF-AM                 & 5.5 (-7.63)      & 3.3 (-9.77)   &  8.8 (-12.11)                      \\
                                    & White                 & 46.9 (23.03)      & 33.6 (-2.39)   &  80.5 (18.41)                \\ \cline{2-5} 
                                    & \textbf{Total:}       & 57.2 (13.24)      & 42.8 (-13.24)  &  100.0                       \\ \hline
\multirow{4}{*}{\textsf{The Wall Street Journal}} & Asian   & 4.9 (-3.91)      & 3.6 (-8.60)   &  8.5 (-9.43)                      \\
                                    & AF-AM                 & 6.1 (-3.41)      & 3.3 (-5.86)   &  9.4 (-6.68)                       \\
                                    & White                 & 51.3 (15.70)      & 30.8 (-3.35)  &  82.2 (12.23)                \\ \cline{2-5} 
                                    & \textbf{Total:}       & 62.3 (10.77)      & 37.7 (-10.77)  &  100.0                       \\ \hline
\multirow{4}{*}{\textsf{BBC News}}  & Asian                 & 5.3 (-2.64)      & 6.6 (-4.49)   & 12.0 (-5.11)                       \\
                                    & AF-AM                 & 7.1 (-1.91)      & 2.7 (-6.01)   &  9.8 (-5.76)                       \\
                                    & White                 & 46.2 (11.00)      & 32.1 (-2.36)   &  78.2 (8.04)                 \\ \cline{2-5} 
                                    & \textbf{Total:}       & 58.6 (7.97)      &  41.4 (-7.97) &  100.0                       \\ \hline
\end{tabular}
\caption{Proportion of news shares by different demographic groups for each news source.}
\vspace{-3mm}
\label{tab:shares_newspaper}
\end{table}

\begin{table}[htbp]
\centering
\frenchspacing
\scriptsize
\begin{tabular}{|c|c|cc|c|}
\hline
\multirow{2}{*}{News media}          & \multirow{2}{*}{Race (\%)} & \multicolumn{2}{c|}{Gender (\%)}        & \multirow{2}{*}{\textbf{Total:}} \\ \cline{3-4}
                                    &                       & \multicolumn{1}{c|}{Male} & Female &                         \\ \hline
\multirow{4}{*}{\textsf{The New York Times}} & Asian        & 12.7 (6.69)       & 10.5 (0.28)     & 23.2 (5.28)    \\ 
                                    & AF-AM                & 11.4 (3.71)       & 3.9 (-4.35)    & 15.2 (-0.36)              \\
                                    & White                 & 35.0 (2.41)        & 26.6 (-6.12)   & 61.5 (-4.08)     \\ \cline{2-5} 
                                    & \textbf{Total:}       & 59.1 (7.97)       & 40.9 (-7.97)   & 100.0                            \\ \hline
\multirow{4}{*}{\textsf{Reuters}}   & Asian                 & 11.3 (5.83)       & 7.9 (-2.97)    & 19.2 (1.71)                       \\
                                    & AF-AM                 & 16.9 (9.97)       & 3.6 (-4.64)    & 20.5 (3.98)                       \\
                                    & White                 & 39.5 (5.74)       & 20.8 (-10.31)   & 60.3 (-4.52)              \\ \cline{2-5} 
                                    & \textbf{Total:}       & 67.7 (15.81)       & 32.3 (-15.81)   & 100.0                        \\ \hline
\multirow{4}{*}{\textsf{The Guardian}} & Asian              & 8.5 (2.22)        & 7.8 (-3.34)    & 16.4 (-1.30)          \\
                                    & AF-AM                 & 10.5 (2.79)       & 3.8 (-4.58)    & 14.3 (-1.04)      \\
                                    & White                 & 41.4 (8.99)       & 27.9 (-5.82)   & 69.3 (1.80)         \\ \cline{2-5} 
                                    & \textbf{Total:}       & 60.4 (10.45)       & 39.6 (-10.45)   & 100.0                        \\ \hline
\multirow{4}{*}{\textsf{The Wall Street Journal}} & Asian   & 9.9 (4.13)        & 8.0 (-3.20)    & 17.9 (0.54)        \\
                                    & AF-AM                 & 14.5 (8.55)       & 4.2 (-4.06)    & 18.8 (3.64)        \\
                                    & White                 & 41.6 (6.97)       & 21.7 (-11.70)   & 63.3 (-3.28)         \\ \cline{2-5} 
                                    & \textbf{Total:}       & 66.0 (13.93)       & 34.0 (-13.93)   & 100.0                        \\ \hline
\multirow{4}{*}{\textsf{BBC News}}  & Asian                 & 12.5 (5.85)       & 11.3 (0.92)     & 23.8 (4.67)                         \\
                                    & AF-AM                 & 12.5 (4.58)       & 2.2 (-6.30)    & 14.7 (-0.59)          \\
                                    & White                 & 34.6 (1.92)        & 26.9 (-5.13)   & 61.5 (-3.25)       \\ \cline{2-5} 
                                    & \textbf{Total:}       & 59.6 (7.57)       & 40.4 (-7.57)   & 100.0                        \\ \hline
\end{tabular}
\caption{Proportion of distinct followers by different demographic groups for each news source.}
\vspace{-3mm}
\label{tab:followers_newspaper}
\end{table}

By comparing between the news sources, we see some obvious patterns: 1) The Wall Street Journal is favored by Male (62.3\%) more than Female (37.7\%); 2) The New York Times has the most balanced gender distribution among spreaders (54.1\% vs 45.9\%); and 3) for The New York Times, The Guardian, and BBC News, the proportion of shares by Asians is greater than by AF-AM.

From a simple comparison to  Table~\ref{tab:demogr_spreaders} which shows the demographic compositions of typical URL sharing behavior, we observed the following trends for all five news sources. First, Males share more news URLs than Female do. Male (54.84\% of news spreaders) issue 54.1\% to 62.3\% of news URL shares. Secondly, Whites share more news URLs than other race groups--White (78.77\% of total users) cover 78.2\% to 82.2\% of news URL shares. 

The Z-values in Table~\ref{tab:shares_newspaper} tell whether the differences between news spreaders and typical URL spreaders are statistically significant or not. The most strong tendency is observed for White-Male. White-Male share more news URLs than they share typical URLs and this tendency is strong (Z $>$ 11\footnote{Z-value is minimum for BBC News, the largest Z-value is 26.24 for The New York Times.}). Then, another observations is that White-Female are less likely to share news URLs than typical URLs (Z $<$ 0) except for The New York Times. On average, White-Male make 46.8\% and White-Female make 32.9\% of news URL shares. From the two proportions, one may think this is because White-Female are less active than White-Male in Twitter. However, our method of comparing the news URL sharing behavior with typical URL sharing behavior can effectively tell that the difference is not because of the activity level, but of the type of URLs. White-Female do share a significant number of typical URLs.




\subsection{Are Spreaders Similar to Followers of Media Sources?}

In the previous analysis, we observed that White-Male are dominant in sharing news URLs. Then, would such pattern find for the Twitter followers of news sources? 


Table \ref{tab:followers_newspaper} presents the demographics of Twitter followers of each news source. Again, the number in the parenthesis is Z-value, reporting how it differs from typical news sharing behavior. Compared to those users who share typical URLs, we observe two main differences of news media followers: 1) there are more male users ($Z>0$); 2) except The Guardian, all the other four news sources have fewer White users ($Z<0$). The New York Times and BBC News have more Asian followers and Reuters and The Wall Street Journal have more Asian and AF-AM users. This results in that the following three groups, Asian-Male, AF-AM-Male, and White-Male, are prominent in the followers of media sources ($Z>0$). In addition, we observe that two news sources, The New York Times and BBC News, have positive Z-values for Asian Female followers. 

For both type of users the followers and the spreaders we observe a ``Male dominant'' pattern, confirming that Male are more interested in news for consumption and spread. However, we find significant differences in demographic compositions between the followers and the spreaders of news. While the followers have a certain degree of racial equality, the spreaders are biased towards one particular race, White. This result is particularly important because so far it was known that individuals affiliated with news media play a large part in breaking the news~\cite{hu2012breaking}. Our observation indicates that breaking news is from not only those followers, but also from these news spreaders who are not necessarily following the news sources in Twitter.



\section{What News Spreaders Share}

We study what news spreaders share along three distinct dimensions: the topical category of news, the demographic trait of the authors (journalist) of a news article, and the linguistic properties of news headlines.


\subsection{By News Category}
We firstly examine which categories of news are shared more by particular demographic groups. To this end, we standardized the names of topical categories for the analysis. For example, we grouped news categories relating to health and life and named ``Health and Life" and grouped news categories relating to science and named ``Science and Tech.". 

\begin{table}[htbp]
\centering
\frenchspacing
\scriptsize
\begin{tabular}{|c|c|cc|c|}
\hline
\multirow{2}{*}{Category}          & \multirow{2}{*}{Race (\%)} & \multicolumn{2}{c|}{Gender (\%)}  & \multirow{2}{*}{\textbf{Total:}} \\ \cline{3-4}
                                    &                       & \multicolumn{1}{c|}{Male} & Female &                         \\ \hline
\multirow{4}{*}{\textsf{World}}     & Asian                 & 4.3 (-6.96)    & 7.8 (-6.32)   &  12.1 (-9.84)               \\ 
                                    & AF-AM                & 6.1 (-6.47)    & 3.2 (-9.35)   &  9.3 (-12.53)                         \\
                                    & White                 & 40.4 (13.71)    & 38.3 (4.62)    &  78.6 (17.67)              \\ \cline{2-5} 
                                    & \textbf{Total:}       & 50.8 (4.55)     & 49.2 (-4.55)   &  100.0                            \\ \hline
\multirow{4}{*}{\textsf{Health and Life}}    & Asian        & 6.8 (-0.18)     & 7.0 (-2.76)   &  13.8 (-2.25)                       \\
                                    & AF-AM                 & 3.3 (-3.82)    & 3.7 (-2.94)   &  7.0 (-5.54)                       \\
                                    & White                 & 41.9 (4.77)    & 37.3 (0.93)    &  79.2 (5.97)                \\ \cline{2-5} 
                                    & \textbf{Total:}       & 52.0 (1.89)     & 48.0 (-1.89)   &  100.0                        \\ \hline
\multirow{4}{*}{\textsf{Science and Tech}}& Asian           & 5.2 (-3.55)    & 5.4 (-6.98)   &  10.5 (-7.61)                     \\
                                    & AF-AM                 & 6.3 (-3.25)    & 1.7 (-10.12)   &  8.0 (-9.17)                        \\
                                    & White                 & 52.6 (19.17)    & 28.8 (-5.95)  &  81.4 (12.34)                \\ \cline{2-5} 
                                    & \textbf{Total:}       & 64.1 (15.74)    & 35.9 (-15.74)  &  100.0                        \\ \hline
\multirow{4}{*}{\textsf{Business}}  & Asian                 & 4.0 (-4.95)    & 5.3 (-6.93)   &  9.3 (-8.13)                     \\
                                    & AF-AM                 & 7.0 (-2.54)    & 3.1 (-5.69)   &  10.0 (-6.30)                        \\
                                    & White                 & 49.4 (15.59)   & 31.3 (-3.48)  &  80.7 (10.45)                \\ \cline{2-5} 
                                    & \textbf{Total:}       & 60.3 (9.76)    & 39.7 (-9.76)  &  100.0                        \\ \hline
\multirow{4}{*}{\textsf{Politics}}  & Asian                 & 5.5 (-3.06)    & 4.6 (-9.52)   &  10.1 (-9.94)                  \\
                                    & AF-AM                 & 6.3 (-4.01)     & 3.2 (-7.53)   &  9.5 (-8.18)                        \\
                                    & White                 & 47.3 (16.91)    & 33.1 (-2.58)   &  80.4 (14.20)                \\ \cline{2-5} 
                                    & \textbf{Total:}       & 59.1 (13.23)    & 40.9 (-13.23)  &  100.0                        \\ \hline
\end{tabular}
\caption{Number of shares by category.}
\vspace{-3mm}
\label{tab:shares_category}
\end{table}

Table~\ref{tab:shares_category} shows the proportion of news shares by each demographic group for each topical category. We consider only topics that were present in all news sources for this analysis. Foremost in Science and Tech, Business, and Politics, we can see the great gender differences. On average, 61.2\% of news URLs of these three topics are shared by Male. In the others two categories, World and Health and Life, Female make more contributions (48.6\% of shares). 

When compared to typical URL sharing behavior, we observe the tendency of White-Male sharing news URLs for all categories ($Z>0$), but the tendency is stronger for Science and Tech, Business, and Politics ($Z>9.76$) than World ($Z=4.55$) and Health and Life ($Z=1.89$). One interesting observation is that White-Female do share more news URLs of World and Health and Life categories than the typical URLs ($Z>0$).

To understand better how demographic traits relate to topical preferences, we compute the relative preferences of each demographic group to ten topical categories (see Figure~\ref{fig:category_top10}). 
News articles about Tech are more likely to be shared by Male than Female. We then see White are more likely to share news about Health and Tech while Asian and AF-AM participate more in sharing news about Sports and Arts. Lastly, Science is favored by Asian but Business is favored by AF-AM. Our analysis shows that demographic groups have different topical tastes in sharing.  This guides us how news media publish their contents to target appropriate user segments.
\subsection{By Author's Demographics}


In this section, we study how the gender of a journalist who wrote a news article influences its shares.  While some differences in topics written~\cite{lynch1993catch} or sources used ~\cite{zeldes2007race} between male and female journalists have been reported~\cite{lynch1993catch} , its appealing to each demographic group has not been  fully explored.

Table~\ref{tab:authors_newspaper} shows the demographics of the authors for each news source. Overall, the proportion of Male authors are higher than that of Female authors--on average, 60.04\% of the authors are Male. Reuters and BBC News have more skewed gender distributions than the other three sources. In terms of race, most of the authors are White (83.8\% on average across five media sources), followed by Asian authors (10.5\%). We observe only 5.7\% of the authors are AF-AM and strikingly low fraction of AF-AM Female authors (1.42\%).



\begin{table}[htbp]
\centering
\footnotesize
\begin{tabular}{|c|c|cc|c|}
\hline
\multirow{2}{*}{News Media}          & \multirow{2}{*}{Race (\%)} & \multicolumn{2}{c|}{Gender (\%)}        & \multirow{2}{*}{\textbf{Total:}} \\ \cline{3-4}
                                    &                       & \multicolumn{1}{c|}{Male} & Female &                         \\ \hline
\multirow{4}{*}{\textsf{The New York Times}} & Asian        & 4.9       & 5.8   &  10.7             \\ 
                                    & AF-AM                 & 3.9      & 0.9  &  4.8                        \\
                                    & White                 & 49.4      & 35.1  &  84.5            \\ \cline{2-5} 
                                    & \textbf{Total:}       & 58.1      & 41.9   &  100.0                            \\ \hline
\multirow{4}{*}{\textsf{Reuters}}   & Asian                 & 6.8        & 6.0    &  12.8                       \\
                                    & AF-AM                 & 4.3       & 2.3  &  6.6                       \\
                                    & White                 & 51.3       & 29.3  &  80.6                 \\ \cline{2-5} 
                                    & \textbf{Total:}       & 62.5       & 37.5   &  100.0                        \\ \hline
\multirow{4}{*}{\textsf{The Guardian}} & Asian              & 3.4      & 4.6    &  8.1               \\
                                    & AF-AM                 & 3.8       & 1.2  &  5.0                       \\
                                    & White                 & 50.4      & 36.6  &  87.0               \\ \cline{2-5} 
                                    & \textbf{Total:}       & 57.6      & 42.4   &  100.0                        \\ \hline
\multirow{4}{*}{\textsf{The Wall Street Journal}} & Asian   & 7.0        & 6.1    &  13.1               \\
                                    & AF-AM                 & 2.9       & 1.6  &  4.5                       \\
                                    & White                 & 47.9      & 34.5 &  82.4                 \\ \cline{2-5} 
                                    & \textbf{Total:}       & 57.8      & 42.2  &  100.0                        \\ \hline
\multirow{4}{*}{\textsf{BBC News}}  & Asian                 & 3.2     & 4.7   &  7.9            \\
                                    & AF-AM                 & 6.3      & 1.1    &  7.4               \\
                                    & White                 & 54.7   & 30.0  &  84.7             \\ \cline{2-5} 
                                    & \textbf{Total:}       & 64.2     & 35.8   &  100.0                        \\ \hline
\end{tabular}
\caption{Demographic characteristics of the collected authors by news source.}
\vspace{-2mm}
\label{tab:authors_newspaper}
\end{table}

Table~\ref{tab:confusion_matrixes} shows the proportion of the spreaders who shared any news URLs written by a certain author demographic group for each news source. 



\begin{table}[]
\centering
\small

\subtable[\large The New York Times]{
\begin{tabular}{ccccccccl}
\cline{4-9}
                                                                                                       &                       & \multicolumn{1}{c|}{}                & \multicolumn{6}{c|}{\textbf{Spreaders (\%)}}                                                                    \\ \cline{4-9} 
                                                                                                       &                       &                                      &                  &                  &                   &                 &                   &                 \\ \cline{4-9} 
                                                                                                       &                       & \multicolumn{1}{c|}{}                & \multicolumn{3}{c|}{Male}                    & \multicolumn{3}{c|}{Female}                    \\ \cline{1-1} \cline{3-9} 
\multicolumn{1}{|c|}{\multirow{7}{*}{\textbf{\begin{tabular}[c]{@{}c@{}}Authors\\ (\%)\end{tabular}}}} & \multicolumn{1}{c|}{} & \multicolumn{1}{c|}{Male} & \multicolumn{3}{c|}{54.8 }               & \multicolumn{3}{c|}{45.2 }                     \\ \cline{3-3}
\multicolumn{1}{|c|}{}                                                                                 & \multicolumn{1}{c|}{} & \multicolumn{1}{c|}{Female}   & \multicolumn{3}{c|}{53.1}               & \multicolumn{3}{c|}{46.9 }                     \\ \cline{3-9} 
\multicolumn{1}{|c|}{}                                                                                 &                       &                                      &                  &                  &                   &                 &                   &                 \\ \cline{4-9} 
\multicolumn{1}{|c|}{}                                                                                 &                       & \multicolumn{1}{c|}{}                & \multicolumn{2}{c|}{Asian} & \multicolumn{2}{c|}{AF-AM} & \multicolumn{2}{c|}{White} \\ \cline{3-9} 
\multicolumn{1}{|c|}{}                                                                                 & \multicolumn{1}{c|}{} & \multicolumn{1}{c|}{Asian}  & \multicolumn{2}{c|}{11.0 }   & \multicolumn{2}{c|}{10.7 }   & \multicolumn{2}{c|}{78.3}   \\ \cline{3-3}
\multicolumn{1}{|c|}{}                                                                                 & \multicolumn{1}{c|}{} & \multicolumn{1}{c|}{AF-AM}  & \multicolumn{2}{c|}{11.3 }   & \multicolumn{2}{c|}{11.6 }   & \multicolumn{2}{c|}{77.1}   \\ \cline{3-3}
\multicolumn{1}{|c|}{}                                                                                 & \multicolumn{1}{c|}{} & \multicolumn{1}{c|}{White}  & \multicolumn{2}{c|}{10.7 }   & \multicolumn{2}{c|}{10.3 }   & \multicolumn{2}{c|}{79.1}   \\ \cline{1-1} \cline{3-9} 
\end{tabular}
\label{tab:confusion_matrix_nytimes}
}


\subtable[\large Reuters]{
\begin{tabular}{ccccccccl}
\cline{4-9}
                                                                                                       &                       & \multicolumn{1}{c|}{}                & \multicolumn{6}{c|}{\textbf{Spreaders (\%)}}                                                                    \\ \cline{4-9} 
                                                                                                       &                       &                                      &                  &                  &                   &                 &                   &                 \\ \cline{4-9} 
                                                                                                       &                       & \multicolumn{1}{c|}{}                & \multicolumn{3}{c|}{Male}                    & \multicolumn{3}{c|}{Female}                    \\ \cline{1-1} \cline{3-9} 
\multicolumn{1}{|c|}{\multirow{7}{*}{\textbf{\begin{tabular}[c]{@{}c@{}}Authors\\ (\%)\end{tabular}}}} & \multicolumn{1}{c|}{} & \multicolumn{1}{c|}{Male} & \multicolumn{3}{c|}{58.7}               & \multicolumn{3}{c|}{41.3}                     \\ \cline{3-3}
\multicolumn{1}{|c|}{}                                                                                 & \multicolumn{1}{c|}{} & \multicolumn{1}{c|}{Female}   & \multicolumn{3}{c|}{57.5}               & \multicolumn{3}{c|}{42.5}                     \\ \cline{3-9} 
\multicolumn{1}{|c|}{}                                                                                 &                       &                                      &                  &                  &                   &                 &                   &                 \\ \cline{4-9} 
\multicolumn{1}{|c|}{}                                                                                 &                       & \multicolumn{1}{c|}{}                & \multicolumn{2}{c|}{Asian} & \multicolumn{2}{c|}{AF-AM} & \multicolumn{2}{c|}{White} \\ \cline{3-9} 
\multicolumn{1}{|c|}{}                                                                                 & \multicolumn{1}{c|}{} & \multicolumn{1}{c|}{Asian}  & \multicolumn{2}{c|}{13.2}   & \multicolumn{2}{c|}{10.4}   & \multicolumn{2}{c|}{76.5}   \\ \cline{3-3}
\multicolumn{1}{|c|}{}                                                                                 & \multicolumn{1}{c|}{} & \multicolumn{1}{c|}{AF-AM}  & \multicolumn{2}{c|}{14.0}   & \multicolumn{2}{c|}{9.6}    & \multicolumn{2}{c|}{76.3}   \\ \cline{3-3}
\multicolumn{1}{|c|}{}                                                                                 & \multicolumn{1}{c|}{} & \multicolumn{1}{c|}{White}  & \multicolumn{2}{c|}{9.5}    & \multicolumn{2}{c|}{10.9}   & \multicolumn{2}{c|}{79.6}   \\ \cline{1-1} \cline{3-9} 
\end{tabular}
\label{tab:confusion_matrix_reuters}
}


\subtable[\large The Guardian]{
\begin{tabular}{ccccccccl}
\cline{4-9}
                                                                                                       &                       & \multicolumn{1}{c|}{}                & \multicolumn{6}{c|}{\textbf{Spreaders (\%)}}                                                                    \\ \cline{4-9} 
                                                                                                       &                       &                                      &                  &                  &                   &                 &                   &                 \\ \cline{4-9} 
                                                                                                       &                       & \multicolumn{1}{c|}{}                & \multicolumn{3}{c|}{Male}                    & \multicolumn{3}{c|}{Female}                    \\ \cline{1-1} \cline{3-9} 
\multicolumn{1}{|c|}{\multirow{7}{*}{\textbf{\begin{tabular}[c]{@{}c@{}}Authors\\ (\%)\end{tabular}}}} & \multicolumn{1}{c|}{} & \multicolumn{1}{c|}{Male} & \multicolumn{3}{c|}{59.7}               & \multicolumn{3}{c|}{40.3}                     \\ \cline{3-3}
\multicolumn{1}{|c|}{}                                                                                 & \multicolumn{1}{c|}{} & \multicolumn{1}{c|}{Female}   & \multicolumn{3}{c|}{54.4}               & \multicolumn{3}{c|}{45.6}                     \\ \cline{3-9} 
\multicolumn{1}{|c|}{}                                                                                 &                       &                                      &                  &                  &                   &                 &                   &                 \\ \cline{4-9} 
\multicolumn{1}{|c|}{}                                                                                 &                       & \multicolumn{1}{c|}{}                & \multicolumn{2}{c|}{Asian} & \multicolumn{2}{c|}{AF-AM} & \multicolumn{2}{c|}{White} \\ \cline{3-9} 
\multicolumn{1}{|c|}{}                                                                                 & \multicolumn{1}{c|}{} & \multicolumn{1}{c|}{Asian}  & \multicolumn{2}{c|}{13.2}   & \multicolumn{2}{c|}{10.0}   & \multicolumn{2}{c|}{76.8}   \\ \cline{3-3}
\multicolumn{1}{|c|}{}                                                                                 & \multicolumn{1}{c|}{} & \multicolumn{1}{c|}{AF-AM}  & \multicolumn{2}{c|}{14.1}   & \multicolumn{2}{c|}{11.1}   & \multicolumn{2}{c|}{74.8}   \\ \cline{3-3}
\multicolumn{1}{|c|}{}                                                                                 & \multicolumn{1}{c|}{} & \multicolumn{1}{c|}{White}  & \multicolumn{2}{c|}{10.5}   & \multicolumn{2}{c|}{9.7}    & \multicolumn{2}{c|}{79.8}   \\ \cline{1-1} \cline{3-9} 
\end{tabular}
\label{tab:confusion_matrix_guardian}
}


\subtable[\large The Wall Street Journal]{
\begin{tabular}{ccccccccl}
\cline{4-9}
                                                                                                       &                       & \multicolumn{1}{c|}{}                & \multicolumn{6}{c|}{\textbf{Spreaders (\%)}}                                                                    \\ \cline{4-9} 
                                                                                                       &                       &                                      &                  &                  &                   &                 &                   &                 \\ \cline{4-9} 
                                                                                                       &                       & \multicolumn{1}{c|}{}                & \multicolumn{3}{c|}{Male}                    & \multicolumn{3}{c|}{Female}                    \\ \cline{1-1} \cline{3-9} 
\multicolumn{1}{|c|}{\multirow{7}{*}{\textbf{\begin{tabular}[c]{@{}c@{}}Authors\\ (\%)\end{tabular}}}} & \multicolumn{1}{c|}{} & \multicolumn{1}{c|}{Male} & \multicolumn{3}{c|}{66.9}               & \multicolumn{3}{c|}{33.1}                     \\ \cline{3-3}
\multicolumn{1}{|c|}{}                                                                                 & \multicolumn{1}{c|}{} & \multicolumn{1}{c|}{Female}   & \multicolumn{3}{c|}{59.6}               & \multicolumn{3}{c|}{40.4}                     \\ \cline{3-9} 
\multicolumn{1}{|c|}{}                                                                                 &                       &                                      &                  &                  &                   &                 &                   &                 \\ \cline{4-9} 
\multicolumn{1}{|c|}{}                                                                                 &                       & \multicolumn{1}{c|}{}                & \multicolumn{2}{c|}{Asian} & \multicolumn{2}{c|}{AF-AM} & \multicolumn{2}{c|}{White} \\ \cline{3-9} 
\multicolumn{1}{|c|}{}                                                                                 & \multicolumn{1}{c|}{} & \multicolumn{1}{c|}{Asian}  & \multicolumn{2}{c|}{12.5}   & \multicolumn{2}{c|}{9.2}    & \multicolumn{2}{c|}{78.4}   \\ \cline{3-3}
\multicolumn{1}{|c|}{}                                                                                 & \multicolumn{1}{c|}{} & \multicolumn{1}{c|}{AF-AM}  & \multicolumn{2}{c|}{12.3}   & \multicolumn{2}{c|}{9.9}    & \multicolumn{2}{c|}{77.8}   \\ \cline{3-3}
\multicolumn{1}{|c|}{}                                                                                 & \multicolumn{1}{c|}{} & \multicolumn{1}{c|}{White}  & \multicolumn{2}{c|}{9.3}    & \multicolumn{2}{c|}{9.8}    & \multicolumn{2}{c|}{80.9}   \\ \cline{1-1} \cline{3-9} 
\end{tabular}
\label{tab:confusion_matrix_wsj}
}


\subtable[\large BBC News]{
\begin{tabular}{ccccccccl}
\cline{4-9}
                                                                                                       &                       & \multicolumn{1}{c|}{}                & \multicolumn{6}{c|}{\textbf{Spreaders (\%)}}                                                                    \\ \cline{4-9} 
                                                                                                       &                       &                                      &                  &                  &                   &                 &                   &                 \\ \cline{4-9} 
                                                                                                       &                       & \multicolumn{1}{c|}{}                & \multicolumn{3}{c|}{Male}                    & \multicolumn{3}{c|}{Female}                    \\ \cline{1-1} \cline{3-9} 
\multicolumn{1}{|c|}{\multirow{7}{*}{\textbf{\begin{tabular}[c]{@{}c@{}}Authors\\ (\%)\end{tabular}}}} & \multicolumn{1}{c|}{} & \multicolumn{1}{c|}{Male} & \multicolumn{3}{c|}{62.4}               & \multicolumn{3}{c|}{37.6}                     \\ \cline{3-3}
\multicolumn{1}{|c|}{}                                                                                 & \multicolumn{1}{c|}{} & \multicolumn{1}{c|}{Female}   & \multicolumn{3}{c|}{50.0}               & \multicolumn{3}{c|}{50.0}                     \\ \cline{3-9} 
\multicolumn{1}{|c|}{}                                                                                 &                       &                                      &                  &                  &                   &                 &                   &                 \\ \cline{4-9} 
\multicolumn{1}{|c|}{}                                                                                 &                       & \multicolumn{1}{c|}{}                & \multicolumn{2}{c|}{Asian} & \multicolumn{2}{c|}{AF-AM} & \multicolumn{2}{c|}{White} \\ \cline{3-9} 
\multicolumn{1}{|c|}{}                                                                                 & \multicolumn{1}{c|}{} & \multicolumn{1}{c|}{Asian}  & \multicolumn{2}{c|}{3.4}    & \multicolumn{2}{c|}{6.9}    & \multicolumn{2}{c|}{89.7}   \\ \cline{3-3}
\multicolumn{1}{|c|}{}                                                                                 & \multicolumn{1}{c|}{} & \multicolumn{1}{c|}{AF-AM}  & \multicolumn{2}{c|}{13.3}   & \multicolumn{2}{c|}{40.0}   & \multicolumn{2}{c|}{46.7}    \\ \cline{3-3}
\multicolumn{1}{|c|}{}                                                                                 & \multicolumn{1}{c|}{} & \multicolumn{1}{c|}{White}  & \multicolumn{2}{c|}{11.3}   & \multicolumn{2}{c|}{9.1}    & \multicolumn{2}{c|}{79.6}   \\ \cline{1-1} \cline{3-9} 
\end{tabular}
\label{tab:confusion_matrix_bbc}
}
\caption{Confusion matrixes for news authors and spreaders by news source.}
\label{tab:confusion_matrixes}
\end{table}

\subsubsection{Author's Gender}
Does the gender of an author affect the spreading behavior? For The New York Times and Reuters, the proportion of Male spreaders is not significantly different ($<$ 2\%) no matter the gender of the author is.  
However, in the rest three others sources, Male tend to share more news URLs written by Male--the difference is 12.4\% for BBC News, 7.4\% for The Wall Street Journal, and 5.3\% for The Guardian. 
While the effect of the gender of the authors on spreading behavior exists, this might be a mere effect of biological differences in topical tastes--Male and Female journalists write only the topics that readers of the same gender are interested in. 

To control the effect of the topics, we use a Chi-square test~\cite{casella2002statistical} to find which topics are written significantly more by Female (or Male) journalists and which topics are significantly more shared by Female (or Male) spreaders. 
Table~\ref{tab:topic_demographics} shows the graphical presentation of the statistically significant results by Chi-square test statistics ($p < 0.05$). In the table, an upward pointing arrow represents a higher tendency in writing or sharing.  
For example, Male authors write news significantly more about Sport and Opinion, and Female authors write about Health. 
There are no topics that authors and spreaders have the same gender differences except for Health. 
Therefore, 
the gender difference in spreading behavior is unlikely  driven by that in journalists' choice of the topics. 
We bring the potential explanation in later section based on linguistic component of news. 

\subsubsection{Author's Race}
Does the race of an author affect the spreading behavior? We observe that the proportion of Asian spreaders are significantly difference across different race of the authors in all news sources except The New York Times. 
For Reuters, The Guardian, and The Wall Street Journal, Asian spreaders are more likely to share news URLs written by Asian or AF-AM authors. Compared to the proportion of shares by Asian (Table~\ref{tab:shares_newspaper}) which are 10.4\%, 10.7\%, and 8.5\% for those three news sources, respectively, the proportion of the news URLs shares written by Asian authors are increased by 26.9\%, 23.4\%, and 47.1\%, respectively. For AF-AM users, we did not find the same pattern. Lastly, BBC News has a strong tendency that AF-AM share extensively news URLs written by AF-AM and Asian. 


Table~\ref{tab:topic_demographics}(b) shows the discriminative topics for each racial group of authors and spreaders. Asian authors are writing more about World and Tech than White. White authors write more opinionated news articles than Asian. For spreaders, Asian and AF-AM share more Sports news than White. News about Arts is favored by Asian more than White. Once again, we do not find any relationship between the topical interests of a certain racial author group and those of a certain racial spreaders group.




 
\begin{table}[t!]
\begin{center}
\small \frenchspacing

\subtable[\large \textsf{Gender}]{
\begin{tabular}{|c|cc|cc|}
\hline
 \multirow{2}{*}{\textsf{Topic}} & \multicolumn{2}{c|}{\textsf{Author}} & \multicolumn{2}{c|}{\textsf{Spreader}}\\
 \cline{2-5}
 &  \multicolumn{1}{c|}{Female} & Male &  \multicolumn{1}{c|}{Female}  & Male \\
 \hline
 Sport & $\downarrow$ & $\uparrow$ & &\\ 
 Opinion & $\downarrow$ & $\uparrow$ & &\\
 Health & $\uparrow$ & $\downarrow$ & $\uparrow$ & $\downarrow$ \\
 Tech & & & $\downarrow$ & $\uparrow$\\
 Business & & & $\downarrow$ & $\uparrow$ \\ 
\hline
\end{tabular}
\label{tab:topic_gender}
}

\subtable[\large \textsf{Race}]{
\begin{tabular}{|c|ccc|ccc|}
\hline
 \multirow{2}{*}{\textsf{Topic}} & \multicolumn{3}{c|}{\textsf{Author}} & \multicolumn{3}{c|}{\textsf{Spreader}}\\
 \cline{2-7}
 &  \multicolumn{1}{c|}{Asian} &  \multicolumn{1}{c|}{AF-AM} & White & \multicolumn{1}{c|}{Asian} &  \multicolumn{1}{c|}{AF-AM} & White\\
 \hline
 World & $\uparrow$ & & $\downarrow$ & & &\\ 
 Tech & $\uparrow$ & & $\downarrow$ & & & \\
 Opinion & $\downarrow$ & &  $\uparrow$ & & & \\
 Sports & & & & $\uparrow$ & $\uparrow$ & $\downarrow$ \\
 Art & & & & $\uparrow$ & & $\downarrow$ \\ 
 \hline
\end{tabular}
\label{tab:topic_race}
}
\end{center}
\caption{Discriminative topics for gender and race groups by authors and spreaders.}
\vspace{-4mm}
\label{tab:topic_demographics}
\end{table}



\subsection{By LIWC Analysis}

Linguistic Inquiry and Word Count (LIWC)~\cite{pennebaker2001linguistic} is a dictionary-based text mining software. Since it has been proposed, it has been widely used for a number of different tasks, including 
sentiment analysis~\cite{Ribeiro2016} and discourse characterization in social media platforms~\cite{correa-2015-anonymityShades}.  Next, we use LIWC to characterize differences in the content shared by different demographic groups. Its latest version, LIWC 2015 (used in this work),  defines about 90 linguistic categories and classifies more than 6,400 words into those categories~\cite{pennebaker2015development}.  For example, the word `cried' falls into the sadness, negative emotion, overall affect, verbs, and past focus categories.  Then, in a given text, the LIWC software finds the occurrence of the words in each category.  The output is the proportion of the words in each category to the total words in the text. 

\begin{table}[hbt]
\begin{center}
\small \frenchspacing
\begin{tabular}{|l|cc|}
\hline
\textsf{LIWC Dimension} & \textsf{Our data} & \textsf{Newman et al. \cite{newman2008gender}}\\
\hline
Pronouns & & \\
$\quad$ First-person singular & M$<$F & M$<$F \\
$\quad$ Third-person & M$<$F & M$<$F \\
\hline
Linguistic dimensions & & \\
$\quad$ Negations & M$<$F & M$<$F \\
\hline
Current concerns & & \\
$\quad$ Money & M$>$F & M$>$F \\
\hline
Biological process & & \\
$\quad$ Ingestion & M$<$F & - \\
\hline
Spoken categories & & \\
$\quad$ Assent & M$<$F & - \\
\hline
Swear words & M$>$F & M$>$F \\
\hline
Female	references & M$<$F & - \\
\hline
\end{tabular}
\end{center}
\caption{LIWC analysis of ours and Newman et al. [30]}
\vspace{-3mm}
\label{tbl:liwc}
\end{table} 

Table~\ref{tbl:liwc} presents the result of LIWC analysis of headlines shared by Male spreaders and Female spreaders.  For comparison purposes, we also show the result of effects of gender on language use~\cite{newman2008gender}. We show only LIWC dimensions that have more than 20\% differences between Male and Female and omit the rest because the number of the whole dimensions is more than 90. 

In our data, we find exactly the same trend as \cite{newman2008gender}: Female share headlines including more first-person singular pronouns, third-person pronouns, negations, words about ingestion (e.g., dish, eat, or pizza), assent (e.g., agree, yes, or ok), and female references (e.g., girl, her, or mom), and Male share headlines including more words about money (e.g., audit, cash, or owe), and swear words (e.g., damn, or shit).  
Considering that \cite{newman2008gender} observed those language usage patterns in the texts Male or Female \textit{write}, finding the same patterns in the texts he or she \textit{shares} is surprising and interesting.
The spreaders are likely to share the news that is aligned with the language usage of their own.  
While many research have focused on attracting more clicks by tweaking headlines, such as including named-entities  in headlines~\cite{kim2016compete}, we show that those studies can be extended to target specific user segments.  

In addition, we find some results that are aligned with some stereotypes of races (e.g. Asian share headlines including more words related to family). However, we omit the result of LIWC by race of the spreaders because there have been no available references for a systematic comparison.

\section{Importance of Spreaders}

Finally, we study the impact of understanding news spreaders in two ways: 1) extended readership by news spreaders and 2) understanding news popularity and demographics of news spreaders.

For the first, we compare the original followers and followers of spreaders by the number and the demographics. That is, we analyzed how spreaders extend news media's readers.  For example, if followers of the The New York Times are usually white male but spreaders of The New York Times URLs have a lot of Asian followers, then, the role of spreaders is really important not only because it increases the number of audience but also because it brings ``different" audiences. The results are shown below in detail.

\subsection{Extended Readership by Spreaders}

Ideally, to study the audience size reached by spreaders that is not reached directly by news sources profiles, we would like to have at our disposal the followers and friends of all users from our dataset. However, the number of followers and friends of these users surpasses a billion users, which is unfeasible to be crawled given our resources. As an attempt to provide evidence that spreaders can largely benefit audience of news papers in social media systems,  Table~\ref{tab:followersVsfollowersSpreaders} contrasts the number of followers of the news media profiles and the sum of the number of followers of the spreaders of each news source. Although these results do not quantify exactly the extent to which spreaders are able to increase the audience size of news sources, it clearly shows that they play a very important role in many news source audiences. For example, the number of followers in our sample of spreaders from NYTimes contains more than double the number of followers of The New York Times.

\begin{table}[t] \centering 
\begin{tabular} {|l|c|c|}
\hline
 News Media  & \#Followers  & \# Followers \cr 
 & (news media) & (spreaders) \cr 
 \hline
\textsf{The New York Times}   &  32,626,611 &  67,458,732 \cr
\textsf{Reuters}    & 15,946,449 & 11,119,453\cr
\textsf{The Guardian}    & 6,154,465 & 21,120,210 \cr 
\textsf{The Wall Street Journal} &  12,563,525 & 6,193,775 \cr
\textsf{BBC News} &  27,871,624 & 4,713,614 \cr \hline
\end{tabular} 
\caption{Total/Real number of followers of the news sources in Twitter and number of followers of the spreaders that shared news of the news source.}\label{tab:followersVsfollowersSpreaders}
\vspace{-2mm}
\end{table}

We move onto demographic of the followers of news spreaders. First, we collected followers from a sample of 25\% of spreaders from our dataset. For this data sample, the average confidence level for the number of the followers of the spreaders is 6111.154 $\pm$ 66396.94, with a confidence interval of 95\%. After that, we analyze the demographic characteristics of the followers of the spreaders. 

Table~\ref{tab:followersofthespreaders_newspaper} shows the demographics of the followers of news spreaders. Compared with the demographics of the followers of news sources (Table~\ref{tab:followers_newspaper}), we observe the increase in the percentage of Female--the average increase is 9\%. Besides that, for race, the percentage of White is higher--the average increase is 16\%.
We tried to test whether this difference in demographics of spreaders' followers and those of the original followers is statistically significant.    
We define the demographic distribution of the audience for each news media as a six-long vector whose element is a proportion of each demographic group (e.g., Male-Asian, Female-Asian, ..., and Female-White), respectively.  With these vectors, we use the Kolmogorov-Smirnov test, which is a widely used statistical test to check whether two distributions are generated from an identical reference distribution.  However, the difference is not statistically significant (for The New York Times, D = 0.5, p-value = 0.1641).  The main reason is that the length of the vector, six, is too short to get statistical evidence. In future work, we will build demographic vectors for multiple snapshots and compute the statistical significance by concatenating those vectors.  


\begin{table}[htbp]
\centering
\footnotesize
\begin{tabular}{|c|c|cc|c|}
\hline
\multirow{2}{*}{News Media}          & \multirow{2}{*}{Race (\%)} & \multicolumn{2}{c|}{Gender (\%)}        & \multirow{2}{*}{\textbf{Total:}} \\ \cline{3-4}
                                    &                       & \multicolumn{1}{c|}{Male} & Female &                         \\ \hline
\multirow{4}{*}{\textsf{The New York Times}} & Asian        & 4.8      & 5.6   &  10.4             \\ 
                                    & AF-AM                 & 6.3      & 4.2  &  10.5                        \\
                                    & White                 & 41.5      & 37.5  &  79.1            \\ \cline{2-5} 
                                    & \textbf{Total:}       & 52.7     & 47.3   &  100.0                            \\ \hline
\multirow{4}{*}{\textsf{Reuters}}   & Asian                 & 4.8        & 5.4    &  10.2                      \\
                                    & AF-AM                 & 6.3       & 4.0  &  10.4                       \\
                                    & White                 & 42.3       & 37.1  &  79.4                \\ \cline{2-5} 
                                    & \textbf{Total:}       & 53.4      & 46.6   &  100.0                        \\ \hline
\multirow{4}{*}{\textsf{The Guardian}} & Asian              & 4.8      & 5.3    &  10.1               \\
                                    & AF-AM                 & 6.1       & 3.8  &  9.9                       \\
                                    & White                 & 42.7      & 37.2  &  80.0               \\ \cline{2-5} 
                                    & \textbf{Total:}       & 53.6      & 46.4   &  100.0                        \\ \hline
\multirow{4}{*}{\textsf{The Wall Street Journal}} & Asian   & 4.8        & 5.3    &  10.1               \\
                                    & AF-AM                 & 6.1       & 3.9  &  10.0                       \\
                                    & White                 & 43.0      & 36.9 &  79.9                 \\ \cline{2-5} 
                                    & \textbf{Total:}       & 54.0      & 46.0  &  100.0                        \\ \hline
\multirow{4}{*}{\textsf{BBC News}}  & Asian                 & 4.8     & 5.3   &  10.1            \\
                                    & AF-AM                 & 6.1      & 3.8    &  9.8               \\
                                    & White                 & 42.9   & 37.2  &  80.1             \\ \cline{2-5} 
                                    & \textbf{Total:}       & 53.7     & 46.3   &  100.0                        \\ \hline
\end{tabular}
\caption{Demographic characteristics of each the followers of the spreaders by news source.}
\label{tab:followersofthespreaders_newspaper}
\end{table}






\subsection{News Popularity and Demographics}

In the previous section, we show that understanding news spreaders is important as they extend the readership of news media. Another important aspect is whether the demographic traits of news spreaders are relating to the popularity of news. To this end, we collect the number clicks
for each news URL using the Bit.ly API\footnote{\url{https://dev.bitly.com/}}. Then, we compare the popularity of news articles shared by different demographic groups to know whether a certain demographic group share news URLs likely to be more popular.

For gender group, we observe that the news items shared by Female are more clicked that those shared by Male. The differences are statistically significant by Kruskal-Wallis H-test ($H=7.719, p<0.005$). For race, the news articles shared by Asians are more clicked ($H=6.659, p<0.005$). The results show that the demographic information of news spreaders can potentially help in predicting the popularity of news articles.










\section{Concluding Discussion}

The increasing diffusion of news in social media systems, associated with the great power provided to users along the dissemination process, are making these platforms a fertile ground for misleading or fake news propagation. 
The growing use of Twitter as a news' channel highlights the
importance of characterizing news spreaders to understand who they are, what they share and their impact. Next, we briefly discuss implications of our main findings and discuss directions we aim to explore next. \\

\noindent
\textbf{Bias on breaking news stories}:
A widely used tool that users use to find breaking news-stories in online social networks is the Trending stories (or topics)~\cite{fb_trends_updated,twitter_trends}.  
Recently, Facebook has been involved in many controversies related to trending stories~\cite{facebook_transparency}. First, Facebook involved human curators as part of its process to identify trending stories. A main criticism was that human curators could bias the final list of stories. Then, Facebook removed the human intervention and followed the popular perception that 
data-driven algorithms would not be biased as they simply
process data. Our results, however, shows the data itself is biased, at least in terms of the demographic groups considered. We show that demographic groups of white and male users tend to share more news in Twitter. Our results also quantify the existing bias on Twitter shares towards specific demographic groups across news categories and other dimensions.  Thus, our work contributes with a new and important perspective to the emerging debate in the community centered around concerns about bias and
transparency of decisions taken by algorithms operating over user-generated data. Finally, we believe that the increasing availability of information about demographics will help the development of systems that promote more diversity and less inequality to users. 
Thus, as a final contribution of our effort, we intend to release our demographic dataset to the research community by the time of publication of this study. \\

\noindent
\textbf{Personalized news recommendations}:
Our analysis shows different user behaviors in terms of news sharing and also highlight demographic differents in terms of user interests. 
Identifying intrinsic characteristics of the users who spread the news in the online world and identifying how users interest across demographics is a key step towards the development of a framework that can promote the  customization of the user experience using social media for news digest. 
We aim at further exploring this topic as part of our future work by investigating the discriminative power of demographic, linguistic, and network features in predicting a user's interest in specific news and news topics.




\section*{Acknowledgments} \label{sec:acknowledgments}


\small{This work was partially supported by the project FAPEMIG-PRONEX-MASWeb, Models, Algorithms and Systems for the Web, process number APQ-01400-1 and grants from CNPq, CAPES, Fapemig, and Humboldt Foundation.}

\small
\bibliographystyle{SIGCHI-Reference-Format}
\bibliography{references}

\end{document}